\documentstyle[12pt]{article}

\def\JHEP #1 #2 #3{{\sl Journ. High. Energy. Phys.} {\bf#1} (#2) #3}

\def\PRL #1 #2 #3{{\sl Phys. Rev. Lett.} {\bf#1} (#2) #3}

\def\NPB #1 #2 #3{{\sl Nucl. Phys.} {\bf B#1} (#2) #3}

\def\NPBP #1 #2 #3{{\sl Nucl. Phys. B (Proc. Suppl.) } {\bf B#1} (#2) #3}

\def\NPBFS #1 #2 #3 #4{{\sl Nucl. Phys.} {\bf B#2} [FS#1] (#3) #4}

\def\CMP #1 #2 #3{{\sl Commun. Math. Phys.} {\bf #1} (#2) #3}

\def\PRD #1 #2 #3{{\sl Phys. Rev.} {\bf D#1} (#2) #3}

\def\PLA #1 #2 #3{{\sl Phys. Lett.} {\bf #1A} (#2) #3}

\def\PLB #1 #2 #3{{\sl Phys. Lett.} {\bf #1B} (#2) #3}

\def\JMP #1 #2 #3{{\sl Journ. Math. Phys.} {\bf #1} (#2) #3}

\def\PTP #1 #2 #3{{\sl Prog. Theor. Phys.} {\bf #1} (#2) #3}

\def\SPTP #1 #2 #3{{\sl Suppl. Prog. Theor. Phys.} {\bf #1} (#2) #3}

\def\AoP #1 #2 #3{{\sl Ann. of Phys.} {\bf #1} (#2) #3}

\def\PNAS #1 #2 #3{{\sl Proc. Natl. Acad. Sci. USA} {\bf #1} (#2) #3}

\def\RMP #1 #2 #3{{\sl Rev. Mod. Phys.} {\bf #1} (#2) #3}

\def\PR #1 #2 #3{{\sl Phys. Reports} {\bf #1} (#2) #3}

\def\AoM #1 #2 #3{{\sl Ann. of Math.} {\bf #1} (#2) #3}

\def\UMN #1 #2 #3{{\sl Usp. Mat. Nauk} {\bf #1} (#2) #3}

\def\FAP #1 #2 #3{{\sl Funkt. Anal. Prilozheniya} {\bf #1} (#2) #3}

\def\FAaIA #1 #2 #3{{\sl Functional Analysis and Its Application} {\bf

#1} (#2) #3}

\def\BAMS #1 #2 #3{{\sl Bull. Am. Math. Soc.} {\bf #1} (#2) #3}

\def\TAMS #1 #2 #3{{\sl Trans. Am. Math. Soc.} {\bf #1} (#2) #3}

\def\InvM #1 #2 #3{{\sl Invent. Math.} {\bf #1} (#2) #3}

\def\LMP #1 #2 #3{{\sl Letters in Math. Phys.} {\bf #1} (#2) #3}

\def\IJMPA #1 #2 #3{{\sl Int. J. Mod. Phys.} {\bf A#1} (#2) #3}

\def\AdM #1 #2 #3{{\sl Advances in Math.} {\bf #1} (#2) #3}

\def\RMaP #1 #2 #3{{\sl Reports on Math. Phys.} {\bf #1} (#2) #3}

\def\IJM #1 #2 #3{{\sl Ill. J. Math.} {\bf #1} (#2) #3}

\def\APP #1 #2 #3{{\sl Acta Phys. Polon.} {\bf #1} (#2) #3}

\def\TMP #1 #2 #3{{\sl Theor. Mat. Phys.} {\bf #1} (#2) #3}

\def\JPA #1 #2 #3{{\sl J. Physics} {\bf A#1} (#2) #3}

\def\JSM #1 #2 #3{{\sl J. Soviet Math.} {\bf #1} (#2) #3}

\def\MPLA #1 #2 #3{{\sl Mod. Phys. Lett.} {\bf A#1} (#2) #3}

\def\JETP #1 #2 #3{{\sl Sov. Phys. JETP} {\bf #1} (#2) #3}

\def\JETPL #1 #2 #3{{\sl  JETP Lett.} {\bf #1} (#2) #3}

\def\PHSA #1 #2 #3{{\sl Physica} {\bf A#1} (#2) #3}

\def\CQG #1 #2 #3{{\sl Class. Quantum Grav.} {\bf #1} (#2) #3}

\def\SJNP #1 #2 #3{{\sl Sov. J. Nucl. Phys.} {\bf #1} (#2) #3}
\def\FdP #1 #2 #3{{\sl Fortschr. Phys.} {\bf #1} (#2) #3}

\def\SSC #1 #2 #3{{\sl Solid  State Commun.} {\bf #1} (#2) #3}

\def\IJMoP #1 #2 #3{{\sl Int. J. Mod. Phys.} {\bf #1} (#2) #3}

\def\PECAY #1 #2 #3{{\sl Phys. Part. Nucl.} {\bf #1} (#2)
#3}

\def\SPECAY #1 #2 #3{{\sl Sov. J. Part. Nucl.} {\bf #1} (#2)
#3}
\def\PTP #1 #2 #3{{\sl Prog. Theor. Phys.} {\bf #1} (#2) #3}

\def\SSP #1 #2 #3{{\sl Sov. Phys. Solid State} {\bf #1} (#2) #3}

\def\LNP #1 #2 #3{{\sl Lecture  Notes in Phys.} {\bf #1} (#2) #3}

\def\IBID #1 #2 #3{{\sl ibid.} {\bf #1} (#2) #3}

\def\a{\alpha}
\def\b{\beta}
\def\g{\gamma}
\def\d{\delta}
\def\e{\varepsilon}
\def\r{\varrho}
\def\k{\kappa}
\def\l{\lambda}
\def\L{\Lambda}
\def\s{\sigma}
\def\S{\Sigma}

\def\th{\theta}
\def\om{\omega}

\def\G{\Gamma}

\def\t{\tau}
\def\vp{\varphi}

\def\g{\gamma}

\begin{document}

\thispagestyle{empty}
\medskip
\begin{center}
{\Large\bf An Inverse Penrose Limit and Supersymmetry Enhancement in the
Presence of Tensor Central Charges}

\vspace{10mm}
 A.A. Zheltukhin$^{\rm a,b}$ and D.V. Uvarov$^{\rm b} $
\end{center}
\begin{center}
$^{a}$ Institute of Theoretical Physics, University of Stockholm , SCFAB,\\
 SE-106 91 Stockholm, Sweden \\
$^{b}$ Kharkov Institute of Physics and Technology, 61108, Kharkov,  Ukraine
\vskip 1.cm
\end{center}
\begin{center}
{\bf Abstract}
\end{center}

\vskip 5.mm
A connection between weak and strong tension limits and their
perturbative
corrections is discussed. New twistor-like models based on  $D=4,\, N=1$
tensionless superstring and superbrane with tensor central charges are
studied.
 The presence of three, two or less preserved fractions of $\k-$symmetry
in the
actions free of the Wess-Zumino terms is shown. A correlation of extra
$\k-$symmetry with the $R-$symmetry is established.
 The equations of the superstring and superbrane models
preserving  $3/4$ supersymmetry are exactly solved.
The general solution for the Goldstone fermion is pure static, but for the
Goldstone bosons it also includes a term describing string/brane motions
along the
fixed directions given by the initial data. These solutions correspond to the
partial spontaneous breaking of the $D=4,\, N=1$ global supersymmetry and
can be
associated with a static closed magnetic Nielsen-Olesen vortex or a
p-dimensional vortex.

\vspace{10mm}

\section {Introduction}

It has recently been shown \cite{bfhp2} that the maximally supersymmetric
$pp-$wave of IIB superstring \cite{bfhp1} can be obtained as a Penrose
limit of
the supersymmetric $AdS_5$x$S_5$ solution. This ten-dimensional solution is
analogous to the maximally supersymmetric solution which preserves 32
supersymmetries of eleven-dimensional supergravity \cite{kg}.
A physical interpretation of the Penrose limit \cite{pr} as corresponding
to a
large tension (equivalently $weak \;  coupling$) limit of $p-$brane was
suggested in \cite{bfhp2}. The proof of this observation was based on the
rescaling tension $T_p \rightarrow \frac {T_p}{\Omega^(p+1)}$, the  metric
tensor of ten-dimensional supergravity and $k-$form-field potentials $C_k$ in
accordance with the prescription  \cite{guv} on the Penrose limit
extension to
the supergravity case. Taking the limit $\Omega^{(p+1)} \rightarrow 0$ had
resulted to the reduction of brane dynamics to an effective one
corresponding to
brane embedded into a limiting space. This space depends on the choice of
null
geodesic used in the perturbative expansion of the brane action fields. As a
result the brane equations  were simplified and exactly solved
for the
IIB superstring \cite{met}. The resulting IIB superstring theory turned
out to
be dual to the  subsector of $N=4$ super Yang-Mills theory resulting from the
large
$N$ limit of $U(N)$ \cite{mal}. It was noted \cite{son} that the exact
solvability property characterizing the $pp-$wave-like geometry can be
broken if
the Penrose limit is applied to nonconformal backgrounds.

Let us remind that a perturbative approach using the rescalings of tension,
coordinates  and fields was  earlier considered in
\cite{zhe},\cite{rozhe}
for a perturbative solution of non-linear string equations in curved
space-time
\cite{vesa}. A generalization of this approach to the case of $Dp-$branes was
considered in \cite{lizz} with a view of studying the limiting regimes of weak
tension and strong coupling \cite{hul}, \cite{liun}, \cite{guli}.

The dimensionless parameter $\e=\g/\a^\prime$ of the rescaled string tension
introduced in \cite{zhe} can be identified with  $1/\Omega^2$ from
\cite{bfhp2}.
In \cite{zhe} the dimensional parameter $\g\;
([\g]=L^2)$ was assumed to be a characteristic scale of the curved background geometry. 
In view of this the
parameter $\e$ appeared in the primary string equations for
$x^M{(\t,\s)}$
$$
\ddot{x}^M- \left({\g \over \a^\prime}\right)^2 x^{\prime\prime M}
+\G^M_{PQ}(x)\left[\dot{x}^P\dot{x}^Q-\left({\g \over \a^\prime }
\right)^2 x^{\prime P}x^{\prime Q} \right]=0 $$
and in one of the primary Virasoro constraints
$$ (\dot{x}^M G_{MN}\dot{x}^N)
+\left({\g\over \a^\prime}\right)^2 (x^{\prime M} G_{MN} x^{\prime N})=0.
$$
Provided that $\g/\a^\prime<<1$ the string equations were  considered as
non-linear equations with the small parameter $\e$ at the highest derivative
with respect to the world-sheet coordinate $\s$. In the zero approximation
these
equations were reduced to the equations of a light-like geodesic surface
$$
 {\cal D}_{\t}\varphi^{M}_{,\t}(\t,\s)=0
$$
describing tensionless string or null string \cite{sch}, \cite{sta},
\cite{kl},\cite{lirss}. The world sheet of null string is a 2-dimensional
null
geodesic manifold, because it consists of a family of the null geodesics
$\vp(\t,\s)$ enumerated by $\s$ and parametrized by the affine parameter $\t$
along each null geodesic. Remind that the generalization to the case of 
null superstrings, null (super)branes and their quantization were studied in
\cite{zhe2},\,\cite{zhero},\,\cite{bazhe}. See also \cite{bozhi}. To
study the
deformation of null world sheet caused by a weak tension inclusion, the
rescaling
of the world-sheet coordinate $\s$
$$
\s=\e\xi
$$
 was done in \cite{zhe} which was matched with the perturbative expansion for the string world
vector
$$
x^M(\t,\s) =\vp^M(\t,0)+\e\psi^M(\t,\xi)+\e^2\chi^M(\t,\xi) + ...
$$
 This procedure is a mathematical one for fixing the restriction to study the
  world-sheet  deformations in the  vicinity of its geodesic line, e.g.
$\vp^M(\t,\s)|_{\s=0}$, representing the congruent family of null geodesics
forming the null-string world sheet. In the first approximation a small
deformation of the null world sheet is described  by the function $\psi^M(\t,\xi)$ defined by the
linear perturbative equation
$$
\left( {\cal D}^{2}_{\t}-\partial^{2}_{\xi}\right)\psi^{M}
- R^{M}_{\ PQL}\varphi^{P}_{,\t}\varphi^{Q}_{,\t}\psi^{L}=0,
 $$
where $R^{M}_{\ PQL}$ is the Riemann-Christoffel tensor, and the constraints
\cite{zhe}, \cite{rozhe}
$$
\left( \varphi_{M,\t}{\cal D}_{\t}\psi^{M}\right)=0, \:\:
\left( \varphi_{M,\t}\psi^{M}\right)=0.
$$
The equation for $\psi^M$ is the general covariant extension of the geodesic
deviation equation \cite{petr} by the oscillatory term
$\partial^{2}_{\xi}\psi^{M}$. The term  $\partial^{2}_{\xi}\psi^{M}$
signals on
the appearance of  tension  and encodes the effect  of the string elastic force
pushing out the null string points from the null geodesics.
 For the class of symmetric spaces characterized by the condition
$$
R_{MPQL}=\kappa (G_{MQ}G_{PL}-G_{ML}G_{PQ})
$$
the equation for $\psi^{M}$ acquires a simple form of the general
covariant wave
equation
$$
\left( {\cal
D}^{2}_{\t}-\partial^{2}_{\xi}\right)\psi^{M}=0.
$$
The equation was reduced to a linear system of modified Bessel
equations for
the de Sitter and Friedmann-Robertson-Walker spaces and it was exactly 
solved \cite{rozhe}. As a result, the string equations in the de Sitter space
can also be quantized in a  first approximation in the rescaled tension
parameter $\e$
or, equivalently, for a large Hubble constant $H \,\,
(\alpha^{\prime}H^{2}\gg1)$, if the parameter $\g$ is identified with
$1/H^2$.

The limit of strong tension $\Omega^2 \rightarrow 0$  \cite{bfhp1}, or
 equivalently  $\e=\g/\a^\prime \rightarrow \infty$, is the inverse limit
to the above-considered one named here as an inverse Penrose limit.
 Thus, it seems to be expedient to apply the Penrose limiting procedure
\cite{pr}
directly inside  the primary string equations and constraints, because they
are correct at any value of the parameter $\e$. It can be achieved by the
introduction of local coordinates in a neighbourhood of a null geodesic
$\vp^M$
belonging to the above considered geodesic null worldsheet.
After the introduction of these local coordinates we must perform 
their rescaling
accompanied with  the transformation of the background metric  
$g_{MN} \rightarrow g_{MN}(1/ \e)$ following the Penrose prescription.
 At last, the  $\lim\limits_{\e\rightarrow \infty}{[\e g_{MN}(1/ \e)]}$
 is  to be considered in the transformed string equations and constraints.
This limit should give the correct description of string dynamics in the
limiting $pp-$wave background complemented by perturbative corrections
deforming
the zero $pp-$wave approximation of the original background.
Then there arises the  possibility to compare the strong and weak perturbative
expansions in the 
vicinity of the null geodesic $\vp^M$. So, one can try to formulate
the transformations connecting these perturbative solutions to get  a new
information on the structure and symmetries of non-perturbative string
solutions.

 Study of this problem is strongly motivated by the case of $Dp-$brane
theory, where a strong coupling limit of the generalized Born-Infeld
action and
the 2-form ${\cal F} = B+ ({\g\over2\pi\a^{\prime}})^{-1}F$ were studied
\cite{lizz}, \cite{zli} and a connection between tensionless strings/branes,
noncommutative geometry \cite{alb}, \cite{codos}, \cite{dohul},
\cite{seiwi} and
NCOS theory \cite{seisut}, \cite{gomms}, \cite{gomss} was observed. 
In  particular,  the electric field  $E$  of tensionless
$Dp$-brane  was shown to
become constant $E=\g/\pi\a^{\prime}$ (if  $B=0$) and  to  coincide with the
critical value of electric field \cite{gukp} given by the NCOS theory. 
Investigation of the correlations between the space-time structure in M
theory
and possible mechanisms to generate string/brane tension deserves to be
continued.

A duality connection between the strong and weak tension limits hints to
a possibility of the supersymmetry enhancement in the models based on
tensionless
superstrings and superbranes.
This problem can surely  be studied in the string/brane model with the $
D=4 \,
N=1$  supersymmetry enlarged by the tensor central charges (TCC) $Z_{mn}$
describing a brane/string contribution into the supersymmetry algebra.
An interesting example of 1/2 supersymmetric configurations encoded by the 
so-enlarged supesymmetry algebra is the domain wall \cite{azgit},\cite{ggt}. 
Similar  domain wall is also created by the gluino condensate in the $SU(n)$
QCD \cite{ds,gs} due to the spontaneous breakdown of discrete chiral symmetry
\cite{vy}.
A wide class of interesting string/brane models considering the TCC
coordinates $z_{m_1m_2...m_p}$ as new independent degrees of freedom associated with the p-form generators $Z_{m_1m_2...m_p}$ was studied in \cite{curt}, \cite{gren}, \cite{sieg}, \cite{bes}, \cite{se}, \cite{hew}, \cite{ruds2}.

The existence of BPS states preserving $1/4, 1/2, 3/4$ of the $ D=4, \,
N=1$ supersymmetry enlarged by the TCC $Z_{mn}$ was  proved in \cite{gght}
using a
model-independent analysis of the enlarged supersymmetry algebra. The
connection of these BPS states with the Jordan algebra of 4x4 real symmetric
matrices and  associated geometric structures was revealed in \cite{gght},
where
the combinations of momentum and domain-wall charges corresponding to the BPS
states were found. Realizations of supersymmetry configurations with enhanced
supersymmetry (such as 3/4 supersymmetry) in string/M-theory were studied in
\cite{gah},\cite{lupo}, \cite{ueno}. The superparticle model with 3/4
supersymmetry in the enlarged space-time was constructed in
\cite{balu},\cite{ruds}. A general class of $pp-$wave solutions preserving
fractions $\nu$, with $1/2 <\nu<1$, of the supersymmetry in the type $IIB$
theory and  M-theory was described in \cite{clupo1}, \cite{clupo2}. A
$pp-$wave of M-theory preserving 3/4 supersymmetry was realized in
\cite{mic}.
In \cite{gah2} the solutions of M-theory with 18, 20, 22 and 24 extra
supersymmetries were presented.
General algebraic and geometric foundations for the enlargement of
superspace by  adding tensor central charge coordinates  connected with the 
free  differential algebra ideas \cite{dafre} were developed in \cite{caib}.

To clear up the dynamical role of TCC coordinates, a new class of string
models
in the $D=4$ space-time enlarged by the six real TCC coordinates $z_{mn}$  was
constructed in \cite{zli} as a natural generalization of the twistor-like
formulations \cite{gz} for the Nambu-Goto and Schild string actions. 
Note that twistor formulation of the action \cite{gz} was given in \cite{il} 
based on a twistor description of null two-surfaces similar to that of \cite{penca}. Twistor approach to supersymmetry was studied in \cite{fer},\cite{shir},\cite{bebcel},\cite{luknow}.
 In  \cite{zli} the twistor-like model of tensionless string minimally
extended by the term linear in the derivatives of TCC coordinates  was shown 
to be  exactly solvable. It was established that the inclusion of $z^{mn}$ lifts the
light-like character of the tensionless string worldsheet and removes the
degeneracy of the worldsheet metric.
Study of the Hamiltonian structure and symmetries of the minimally extended
model  shows that the string constraints reduce the number of
independent TCC
coordinates $z_{mn}$ to one real effective coordinate  \cite{zli2}.
 This corresponds to the string moving in an effective $(4+1)-$dimensional
target space instead of the original  $(4+6)-$dimensional space-time. The P.B.
algebra for the first-class constraints of the model has a structure
similar to
that of the contracted algebra of rotations of an (Anti) de Sitter space
\cite{hakas}, but the second-class constraints have to deform the P.B.
algebra
into the Dirac bracket algebra. The Lorentz-covariant antisymmetric 9x9
complex
Dirac $\hat{\mathbf C}$-matrix of the P.B. of the second-class constraints
has a
rather  clear algebraic structure  \cite{zli2},   and the  D.B. algebra
construction of the   first-class constraints will shed light on the
structure
of the effective 5-dimensional target-space connected with the TCC. It seems
reliable to connect this space with the appropriate  boundary of the
convex cone
analysed in \cite{gght}.

An effective role of TCC in the transition from ten-dimensional Minkowski
space-time to $AdS_5$x$S_5$ was observed in \cite{pol}. The
connections
between TCC, symplectic superalgebras and the theory of massless higher spin
fields and currents were discussed in \cite{vas}.
So, one may suppose that the twistor-like actions \cite{gz,zli2},
constructed from
very simple geometric objects in the space-time enlarged by TCC
coordinates and
auxiliary twistor-like variables can code a hidden dynamics relevant to that
 in the backgrounds with an enhanced supersymmetry.

In this paper we study supersymmetric generalizations of the twistor-like
string
models with TCC \cite{gz,zli2} and propose new models in the enlarged
superspace
which preserve $3/4,1/2$ or smaller fractions  of $D=4 \, N=1$ global
supersymmetry. A basic  role of the presence of  TCC coordinates for the
supersymmetry enhancement is clarified. We show that the superstring and
superbrane models preserving $3/4$ supersymmetry are exactly solvable, and
find
their general solutions which
correspond to partially broken $D=4,\, N=1$  global supesymmetry.
The general solution for the Goldstone fermion \cite{vol},\cite{volak},
\cite{hupol},\cite{hulpol} is a pure static one. The solution for the
Goldstone
bosons includes a pure static term  and a term describing string/brane
motions
along fixed directions given by the initial data of the considered
variational problem.
These solutions  can be associated with a static closed spin or magnetic
Nielsen-Olesen vortex \cite{no}.
 We extend these results to the case of some exactly solvable super p-brane
models generalizing the models of tensionless p-branes and associate their
Goldstone solutions with p-dimensional magnetic vortices.

In Sect. 2 we formulate the transformation rules for the coordinates of the
enlarged superspace with respect to the global supersymmetry and $\k-$symmetry.
The  enlarged 4x4 matrix of the Cartan one-forms $W_{ab}$ symmetric in the
Majorana spinor indices  $a,b$ is presented and its $\k-$symmetry
transformation
rules are defined.

In Sect. 3 the matrix elements of the matrix one-form $W_{ab}$ in the linear
space of local Newman-Penrose dyads are studied, and the conditions for their
invariance under the $\k-$symmetry are formulated. We show that the maximum
supersymmetry enhancement, preserving $3/4$ supersymmetry, takes place for
the
diagonal matrix elements only.

In Sect. 4 we connect  the  maximum  enhancement of the supersymmetry with the
presence of the $R-$symmetry in the diagonal Cartan one-forms.

In Sect. 5  these scalar diagonal one-forms are used to construct a new model of
the  $D=4, \, N=1$ superstring preserving $3/4$ supersymmetry. The exact
solutions
of the model are presented there.

In Sect. 6 the above-discussed results are generalized to the case of super
p-branes, and their actions preserving  $3/4$ supersymmetry are presented. We
show that the general solutions for the Goldstone fermion and bosons are
similar
to the superstring ones, but additional constraints for the initial data
appear
here.

Finally, Sect. 7 contains conclusions and discussion concerning  possible ways of application of the obtained results.

\section {Supersymmetry, $\kappa$-symmetry and \\the Newman-Penrose dyads}

    It is well known that the inclusion of the tensor central charge (TCC)
2-form $Z_{mn}$\\ modifies the anticommutator of the  $D=4$,  $ N=1$  Majorana
supercharges $Q_a$
\begin{equation}\label{1}
\{ Q_a ,Q_b \}=(\g^m\,C^{-1})_{ab}P_{m}+i(\g^{mn}\,C^{-1})_{ab}Z_{mn}.
\end{equation}
The real parameters  $z_{mn}$ corresponding to $Z_{mn}$ may be presented
in the
spinor form
\begin{equation}\label{2}
z_{ab}=iz_{mn}(\g^{mn}\,C^{-1})_{ab},
\end{equation}
by analogy with the spinor representation for the space-time coordinates
$x_{m}$
\begin{equation}\label{3}
x_{ab}=x_{m}(\g^{m}\,C^{-1})_{ab}.
\end{equation}

The coordinates $z_{ab}$ and $x_{ab}$ may be treated as the components of
the real symmetric spin-tensor
\begin{equation}\label{4}
Y_{ab}\equiv
x_{ab} + z_{ab}
\end{equation}
associated with the 4x4  spinor matrix $Y_a{}^b$
\begin{equation}\label{5}
Y_a{}^b=Y_{ad}C^{db} =\left(\begin{array}{cc} z_{\a}{}^{\b}&x_{\a\dot\b }\\
\tilde x^{\dot\a \b}&\bar z^{\dot\a}{}_{\dot\b}\\
\end{array}\right), \quad  z_{\a}{}^\b=\e_{\a\g} z^{\g\b},
\end{equation}
where $C$ is the charge conjugation matrix
\begin{equation}\label{6}
C^{ab}=\left(\begin{array}{cc} \e^{\a\b}&0\\0&\e_{\dot\a\dot\b}\\
\end{array}\right)
\end{equation}
chosen to be imaginary in the Majorana representation similarly to the
$\g^m-$matrices.
The matrix $Y_a{}^b$ (\ref{5}) paired with the Grassmann Majorana spinor 
$\th_a
$
\begin{equation}\label{7}
\th_a=\left(\th_\a\atop \bar\th^{\dot \a}\right) ,\quad
\th^a=C^{ab}\th_b, \quad
 \th^{\a} =\e^{\a\b}\th_\b
\end{equation}
is a compact generalized coordinate convenient to construct superstring
models
in the enlarged superspace $(Y_{ab}, \th_a)$, because the $D=4\; N=1$ global
supersymmetry  transforms the superspace $(Y_{ab}, \th_a)$ into itself
 \begin{equation}\label{8}
\d_\e\th_a=\e_a ,\quad \d_\e Y_{ab}=2i(\th_a\e_b +\th_b\e_a ).
\end{equation}

The Cartan differential one-forms $W_a$ and  $W_{ab}$ invariant under the
transformations (\ref{8}) are given by
\begin{equation}\label{9}
W_a=d\th_a,\quad W_{ab}=dY_{ab}-2i(d\th_a\th_b + d\th_b\th_a ).
\end{equation}
A realization of $\kappa-$symmetry transformations may be chosen in the form
\begin{equation}\label{10}
\d_\k\th_a=\k_a  ,\quad
\d_\k Y_{ab}=-2i(\th_a\k_b + \th_b\k_a ),
\end{equation}
where $\k_a(\t,\s)$ is a Majorana spinor.
Using (\ref{9}) and (\ref{10}) we find the transformation rules for the
supersymmetric Cartan one-forms under the $\kappa-$symmetry
\begin{equation}\label{11}
\d_\k W_{a}=d\k_a, \quad
\d_\k W_{ab}=-4i(d\th_a\k_b + d\th_b\k_a ).
\end{equation}

Following the approach studied in \cite{zli} we intend here to extend the
$(Y_{ab},\th_a)$-superspace by the addition of the local Newman-Penrose dyads
$u^\a$, $v_\a$ attached to the superstring  worldsheet. The dyads are suitable
twistor-like variables  defined by their properties
\begin{equation}\label{12}
u^\a v{_\a} \equiv u^\a\e_{\a\b}v^\b=1 ,\quad u^{\a}u_{\a}=v^{\a}v_{\a}=0
\end{equation}
and may be treated as  the  Weyl components of the Majorana spinors $ U_a$
and
$ V_a$
\begin{equation}\label{13}
 U_a=\left(u_\a\atop \bar u^{\dot \a }\right),\,V_a=\left(v_\a\atop \bar
v^{\dot
\a }\right),\quad (U^a \g_{5a}{}^b V_b)=-2i,
\end{equation}
where $(\g_5)_a{}^b$  is the diagonal matrix in the Weyl representation
\begin{equation}\label{14}
(\g_5)_a{}^b=\left(\begin{array}{cc}-i\d_{\a}{}^{\b}&0\\0&i\d^{\dot\a}{}_{\dot\b}\\
\end{array}\right).
\end{equation}
Applying the ideas advanced in \cite{vz} (see also \cite{vst},\cite{vstz}) we
assume here that the dyads $U_a$ and $V_a$ are invariants of the
transformations (\ref{8}) and (\ref{10})
\begin{equation}\label{15}
\d_\e U_{a}=\d_\e V_{a}=0, \quad  \d_\k U_{a}=\d_\k V_a=0.
\end{equation}

Having fixed the transformation rules for the  fields  $Y_{ab}$, $\th_a$ ,
 $U_a$ and $V_a$ one can proceed to the construction of new
$\kappa-$invariant
actions for superstrings with the TCC coordinates.

\section{$\k$-symmetry and invariant one-forms}

   Let us consider the one-form $W_{ab}$ (\ref{9}) and construct the following
supersymmetric and  Lorentz invariant set of one-forms
\begin{equation}\label{16}
 W_{\vp\chi}^{(IJ)}=\vp_a(IWJ)^{ab}\chi_b,
\end{equation}
where  the Majorana spinors $\vp_{a}$ and $\chi_a$ are the linear combinations of
$U_a$, $V_a$ and $(\g{_5}U)_a $, $(\g{_5}V)_a $ forming the 
representation of
the dyad space (\ref{12})-(\ref{13}) by the Majorana spinors
\begin{eqnarray}\label{17}
\vp_{a}=[(\vp^{(u)}_{R}- \vp^{(u)}_{I}\g{_5})U
+ (\vp^{(v)}_{R}-\vp^{(v)}_{I}\g{_5})V]_{a},\nonumber\\
\chi_{a}=[(\chi^{(u)}_{R}- \chi^{(u)}_{I}\g{_5})U
+ (\chi^{(v)}_{R}-\chi^{(v)}_{I}\g{_5})V]_{a}
\end{eqnarray}
and $I^{ab}$, $J^{ab}$  are presented by  the antisymmetric $\g-$matrices
\begin{equation}\label{18}
I^{ab}, J^{ab} = \{ C^{ab},\,(C\g_5)^{ab},\,p^{m}(C\g_m\g_5)^{ab} \},
\end{equation}
where $p^{m}$ is a $\kappa-$invariant vector or pseudovector in the $D=4$
space-time
\begin{equation}\label{19}
p_{m}=(\tilde\vp C\g_m\g_5 \tilde\chi), \;\, or \,\;  (\tilde\vp C\g_m
\tilde\chi)
\end{equation}
constructed of the Majorana spinors $\tilde\vp $, $\tilde\chi$ selected
from the
dyad space (\ref{17}) invariant under the $\k-$symmetry (\ref{15}).
The analysis of the case when  $I^{ab}$, $J^{ab}$ are symmetric matrices
\begin{equation}\label{20}
 I^{ab},
J^{ab}=\{\pi^{m}(C\g_m)^{ab},\,\pi^{mn}(C\g_{mn})^{ab}\}\end{equation}
just as the case of  $I$ and $J$ belonging to the various  sets is reduced to
the analysis of the case (\ref{18}).
 Thereat  using (\ref{11}) and (\ref{15}) we find the
$\k-$transformation
 of $ W_{\vp\chi}^{(IJ)}$
\begin{equation}\label{21}
\d_\k W_{\vp\chi}^{(IJ)}=-4i[(\vp I d\th)(\k J \chi) -(\vp I \k)(d\th J
\chi)]
\end{equation}
and conclude that the  $\k$-invariance of $W_{\vp\chi}^{(IJ)}$
allows some off-shell restrictions for the $\kappa-$symmetry parameter
$\k_a$.
 Note the change of the sign inside the square brackets (\ref{21}) for the
case of $I$ and $J$ belonging to the various matrix sets. For the
antisymmsetric
set  (\ref{18}) these restrictions are subdivided into three cases
$A, B$ and $C$.

For the  $A$-case $I^{ab}$ and  $J^{ab}$ take their values from the shortened 
subset (\ref{18}) 
\begin{equation}\label{22}
I^\prime, J^\prime=(C,\, C\g_5)
\end{equation}
and we obtain  the following two real off-shell conditions for the four real
components $\k_a$
\begin{equation}\label{23}
(\k J^{\prime} \chi)=0, \; \;(\k I^{\prime}\vp)=0
\end{equation}
providing  partial $\k$-invariance of $W_{\vp\chi}^{(I^{\prime}
J^{\prime})}$
under 1/2 fraction of the full $\k$-symmetry (\ref{10}).

The $B-$case corresponds to the choice
\begin{equation}\label{24}
I=I^{\prime}=(C,\, C\g_5),\; \; J=p^m(C\g_5 \g_m)
\end{equation}
or vice versa. It is easy to see that any matrix $\G_A$ from the complete
set of
the sixteen $\g-$matrices when  multiplied by $p_A$ maps the dyad space onto
itself
\begin{equation}\label{25}
(p^A \G_A) \vp=\tilde\vp,\;\; p^A \equiv(\l\G^A\psi)
\end{equation}
 including the case $\tilde\vp$=0. This property of the dyad space 
follows from
the completeness of the Pauli matrices
$\s_{m\a\dot\a}\tilde\s^{m\dot\b\b}=-2\d_{\a}^{\b}\,\d_{\dot\a}^{\dot\b}$ and
the contraction conditions (\ref{12}).  Using this observation we find that
 the B-case is reduced to the A-case yielding two real conditions for the
components of $\k_a$.  The same conclusion remains valid for the $C-$case
\begin{equation}\label{26}
I= p^m(C\g_5 \g_m),\; \, J=\tilde p^m (C\g_5 \g_m),
\end{equation}
corresponding to the one-form $ W_{\vp\chi}^{(C\g_5 p^m\g_m,\:C\g_5 \tilde
p^m\g_m )}$. We conclude  that all the one-forms (\ref{16}) containing different
spinors from the dyad space (\ref{17}) are invariant under 1/2 
$\k-$sym\-met\-ry.
Therefore, without loss of generality one can be restricted to 
studying the
invariant one-forms which belong to the set (\ref{22}).

Now  we observe the certain case of the symmetry enhancement when one extra
$\k-$symmetry appears. This   case corresponds to the choice
$\vp=\chi$ and $I=J$, or equivalently to
\begin{equation}\label{27}
I^{\prime}=J^{\prime}=(C,\,\; C\g_5),\,\:\;  \vp=\chi,
\end{equation}
because then  both  of the  conditions (\ref{23}) coincide giving one
real condition
\begin{equation}\label{28}
(\k J^\prime \vp)=0.
\end{equation}
As a result, each of the one-forms $W_{\vp}$
\begin{equation}\label{29}
W_{\vp}\equiv W_{\vp\vp}^{( C, C)}=\vp_{a}W^{ab}\vp_{b},\;\;\,
 (\k C \vp)=0
\end{equation}
and $W_{\g_5 \vp}$
\begin{equation}\label{30}
W_{\g_5 \vp}\equiv W_{\vp \vp}^{( C \g_5,\, C \g_5)}=(\g_5 \vp)_{a}W^{ab}
(\g_5 \vp)_{b}, \,\;\;
(\k C \g_5 \vp)=0
\end{equation}
becomes invariant under the corresponding three parametric $\k-$symmetry.
Thus,
we arrive to the one-forms preserving $3/4$  $\k-$symmetry in the superspace
$(Y_{ab}, \th_a, u_\a, v_\a)$.

To clarify  this result we note that the $k-$variation ($\ref{21})$ takes the
form
\begin{equation}\label{31}
\d_\k W_{\vp}|_{\om_{\a\b}=0}=4i[(\vp^\a d\th_\a)(\bar \k_{\dot \b} \bar
\vp^{\dot \b} )-( \k_{\b} \vp^{\b})(\bar \vp^{\dot \a} d\bar\th_{\dot \a})]
\end{equation}
in the absence of the TCC coordinates, as it follows from the component
representation
\begin{equation}\label{32}
 W_{\vp}=2\vp^\a \om_{\a\dot\a}\bar \vp^{\dot \a}+\vp^\a \om_{\a}{}^{\b}
\vp_{\b}+ \bar \vp_{\dot\a}\bar \om^{\dot\a}{}
_{\dot\b}\bar \vp^{\dot\b}.
\end{equation}
Eq. (\ref{31}) shows that  $\k-$invariance of
$W_{\vp}|_{\om_{\a\b}=0}$ implies  two real off-shell conditions
\begin{equation}\label{33}
Im(\k_\a \vp^\a)=0,\,\,  Re(\k_\a \vp^\a)=0
\end{equation}
instead of the one real condition (\ref{28})
\begin{equation}\label{34}
Im(\k_\a \vp^\a)=0.
\end{equation}
So, we reveal the cancellation of the contributions into the real part of
$(\k_\a \vp^\a)$ given by the one-form
$\om_{\a\dot\a}$ connected with $x_m$ and the one-forms $\om_{\a}{}^{\b}$,
$\bar \om^{\dot\a}{}_{\dot\b}$ connected with the TCC coordinates $z_{mn}$.
Analogously, the $\k-$invariance of $W_{\vp}$ in the absence of
the space-time coordinates $x^m$ implies the same restrictions ($\ref{33})$
  due to the relation
\begin{equation}\label{35}
\d_\k W_{\vp}|_{\om_{\a\dot\b}=0}=-8i[(\vp^\a d\th_\a)( \k_{\b} \vp^{\b})-
(\bar \k_{\dot \b} \bar \vp^{\dot \b})(\bar \vp^{\dot \a} d\bar\th_{\dot
\a})].
\end{equation}
 Next  we observe that the opposite case of the total $\k-$symmetry breakdown
may be realized by consideration of a linear combination of the
primary  and
partially $\k-$invariant one-forms (\ref{16}) or by using $p_A$ which is not 
a $\k-$invariant object.

Therefore, we see that the appearance of the third extra $\k-$symmetry is a
collective effect of the addition of  $z_{mn}$ and of only one spinor from
the dyad
space.
 This result can be understood from the point of view of symmetry, because the
third
 $\k-$symmetry is accompanied with the  $R$-symmetry.
We shall discuss this observation in the  next section.

\section{3/4 $\k-$symmetry and the $R$-symmetry}

The $R$-symmetry is defined by the $U(1)$ transformations of the Weyl
$\th-$spinor
\begin{equation}\label{36}
\th^\prime_{\a}=e^{-ia_R} \th_{\a}, \, \,
 \bar\th^\prime_{\dot\a}=e^{ia_R}\bar\th_{\dot\a},
\end{equation}
where $a_{R}$ is a real parameter. Alternatively, (\ref{36}) may be presented as an axial
rotation
\begin{equation}\label{37}
\th^\prime_a
=\left(e^{-ia_R}\th_\a\atop e^{ia_R}\bar\th^{\dot\a}\right)
=(e^{a_R \g_{5}}\th)_a
\end{equation}
of the Majorana bispinor $\th_a$ using the relation $e^{a_R
\g_{5}}=cosa_{R}+\g_{5} sina_{R}$.

The anticommutation relation (\ref{1})
\begin{equation}\label{38}
\{ Q_\a ,Q_\b \}=Z_{\a\b}, \, \,\,
\{\bar Q_{\dot\a} ,\bar Q_{\dot\b}\}=\bar Z_{\dot\a\dot\b}
\end{equation}
 fixes the transformation rules for TCC coordinates under the $R-$symmetry
(\ref{36})
\begin{equation}\label{39}
z^\prime_{\a\b}=e^{-2ia_R}z_{\a\b}, \, \,\,
\bar z^{\prime}_{\dot\a\dot\b}=e^{2ia_R}\bar z_{\dot\a\dot\b}.
\end{equation}
In the Majorana representation the axial rotation (\ref{37}) can be rewritten
as
\begin{equation}\label{40}
z^\prime_{a}{}^{b}=
(e^{a_R \g_5}ze^{a_R \g_5})_{a}{}^{b}.
\end{equation}
As a consequence of (\ref{37}) and (\ref{40}),  we find the
$R-$transformations of
$Y_{a}{}^{b}$ (\ref{5}) and $W_{a}{}^{b}$  (\ref{9})
\begin{eqnarray}\label{41}
Y^\prime_{a}{}^{b}=(e^{a_R \g_5}Ye^{a_R \g_5})_{a}{}^{b}, \nonumber  \\
W^\prime_{a}{}^{b}=(e^{a_R \g_5}We^{a_R \g_5})_{a}{}^{b}.
\end{eqnarray}
As a result of (\ref{41}), the $R-$transformation of the one-form $W_\vp$ is
given by
\begin{equation}\label{42}
W^\prime_{\vp}\equiv (\vp W^\prime \vp)=(\vp^\prime W \vp^\prime),
 \end{equation}
where the transformed spinor $\vp^\prime$ is defined as
\begin{equation}\label{43}
\vp^\prime= e^{a_R \g_5} \vp.
\end{equation}
Right now we observe that the $\g_5-$transformation (\ref{43}) belongs to the
local  group \\$ [U(1)$x $O(1,1)]_{dyad} $ transforming the base dyad 
spinors $
U_a$ and  $ V_a$
\begin{eqnarray}\label{44}
U^\prime=e^{-(\a_I + \a_R \g_5)}U, \nonumber \\
 V^\prime=e^{(\a_I + \a_R \g_5)}V.
\end{eqnarray}
The gauge group  $[U(1)$x$ O(1,1)]_{dyad} $ is  the symmetry group of the
relations (\ref{12}) which defines  the  Newman-Penrose dyads
\begin{equation}\label{45}
u^\prime_{\a}=e^{i\a}u_\a, \: v^\prime_{\a}=e^{-i\a}v_\a,
\end{equation}
where $\a$ is a  complex parameter
\begin{equation}\label{46}
\a=\a_R + i\a_I .
\end{equation}

Without loss of generality one can choose  $\vp$ in (\ref{42}) and (\ref{43})
to be equal to  $U$  (or  $V$), thus  making  evident the fact  that the  axial
rotation (\ref{43}) is compensated by the $[U(1$)x $O(1,1)]_{dyad}$
transformations (\ref{44}) defined by
\begin{equation}\label{47}
\a_R =a_R , \: \a_I=0.
\end{equation}
Due to this compensation we may consider the one-form $W_U$ (\ref{29}) as an
effective invariant  of the $R-$symmetry.
On the other hand, the analogous  compensating  $[U(1)$x $O(1,1)]_{dyad} $
transformation for the  one form  $W_V$ (\ref{29}) is defined  by the
conditions
\begin{equation}\label{48}
\a_R =-a_R , \: \a_I=0
\end{equation}
as it follows from (\ref{43}).
We see, therefore, that  the conditions  (\ref{47}) and  (\ref{48}) are
mutually
exclusive,  and   $W_V$ breaks  the $R-$symmetry if $W_U$ preserves it.

In addition we observe that the compensating transformation
(\ref{47}) also allows to consider the product  $(\k_a U^a)$ as an effective
invariant of the $R-$symmetry and the constraint (\ref{28})
\begin{equation}\label{49}
(\k_a \vp^a)|_{\vp=U}=0
\end{equation}
does not break the $R-$symmetry. In that case, however, the product $(\k_a
V^a)$
is not invariant of the $R-$symmetry and the condition (\ref{28})
\begin{equation}\label{50}
(\k_a \vp^a)|_{\vp=V}=0
\end{equation}
implies the $R-$symmetry breaking. We see that the conditions of the  presence
of the $R-$symmetry are the same as those for the presence of extra $\k-$symmetry
and the
mutual presence of the conditions  (\ref{49}) and (\ref{50}) breaks this
extra
$\k-$symmetry.

Thus, we have fixed the correlation between the extra $\k-$symmetry and the
$R-$symmetry.
The one-form $W_\vp$ invariant under $3/4$ $\k-$symmetry and the $R-$sym\-met\-ry
 will be used to build a superstring action preserving $3/4$ supersymmetry.

\section{A superstring model with $3/4$ $k-$symmetry}

An example of the superstring action with extra $\k-$symmetry may be
formulated
in terms of the invariant one-form $W_\vp$ discussed in the  previous section
\begin{equation}\label{51}
S_\vp=\frac{k}{2}\int d\t d\s \r^\mu (\vp_a W_\mu^{ab} \vp_b),
\end{equation}
where $ W_\mu^{ab}$ is the world-sheet pullback of the one-form  $W^{ab}$
\begin{equation}\label{52}
 W_\mu^{ab}\equiv \partial_\mu Y^{ab}+ 2i(\partial_\mu \th^a\th^b+
\partial_\mu \th^b\th^a), \;\;(\mu=(\t,\s))
\end{equation}
and $\r^\mu(\t,\s)$ is a  $\k-$invariant world-sheet density which provides
the
reparametrization invariance of $S_\vp$ analogously to the cases of
tensionless
superbrane \cite{bazhe} or the Green-Schwarz superstring \cite{bz2}.  The
dimensional constant $k$ in (\ref{52}) can be moved  in a redefinition of $x_m$
and $z_{mn}$ making all dynamical variables dimensionless. We consider a closed
superstring and fix $\vp_a$
\begin{equation}\label{53}
\vp_a=U_a.
\end{equation}
 Then the action (\ref{52}) transforms into the action
\begin{equation}\label{54}
S_U=\frac{1}{2}\int d\t d\s \r^\mu (U_a W_\mu^{ab} U_b)
\end{equation}
invariant (by construction) under the three parametric $\k-$symmetry
(\ref{10})
\begin{equation}\label{55}
 \d_\k \r^\mu=0,  \:\, (\k_a U^a)=0.
\end{equation}
  The variations of $S_U$ with respect to $\r_\mu$, $U{^b}$,  $Y_{ab}$ and
$\th_b$ give the equations of motion
\begin{equation}\label{56}
U^a[\partial_\mu Y_{ab}- 2i(\partial_\mu \th_a\th_b
+ \partial_\mu \th_b\th_a)]U{^b}=0,
\end{equation}
\begin{equation}\label{57}
 \r^\mu [\partial_\mu Y_{ab}- 2i(\partial_\mu \th_a\th_b
+ \partial_\mu \th_b\th_a)]U{^b}=0,
\end{equation}
\begin{equation}\label{58}
\partial_\mu (\r^\mu U^a U{^b})=0,
\end{equation}
\begin{equation}\label{59}
(\r^\mu\partial_\mu \th_b)U{^b}=0.
\end{equation}

In view of $3/4$ $\k-$symmetry of $S_U$ we can fix its  gauge by the
conditions
\begin{equation}\label{60}
(V^a \th_a)=0,\;\,(V^a \g_{5a}{}^b\th_b)=0,\;\,(U^a \g_{5a}{}^b\th_b)=0
\end{equation}
and introduce the  $\k-$invariant Grassmann variable $\eta$
\begin{equation}\label{61}
\eta\equiv\frac{1}{2i}(U^a \th_a)
\end{equation}
which incodes the rest of the dynamical degrees of freedom  $\th^a$  and
describes the Goldstone fermion. Next let us take into account the dyad
expansion of  $\th_\a$
\begin{eqnarray}\label{62}
\th_{\a}=\th^{(u)}u_{\a} +\th^{(v)}v_\a,
\nonumber \\
\bar\th_{\dot\a}=\bar\th^{(u)}\bar u_{\dot\a}+\bar\th^{(v)}\bar v_{\dot\a}
\end{eqnarray}
 which may be presented in terms of the Majorana spinors as
\begin{equation}\label{63}
\th_{a}=[(\th^{(u)}_{R}- \th^{(u)}_{I}\g{_5})U + (\th^{(v)}_{R}-
 \th^{(v)}_{I}\g{_5})V]_{a}.
\end{equation}
With the help of (\ref{63}) we find the following  representation
\begin{equation}\label{64}
\th_{a}=-\th^{(v)}_{I}(\g{_5}V)_a=-\eta(\g{_5}V)_a.
\end{equation}
for $\th_\a $ as a function of $\eta$.
To solve Eqs. (\ref{56})-(\ref{59}) we choose an additional gauge fixing
\begin{equation}\label{65}
\r^\s(\t,\s)=0
\end{equation}
 allowed by the reparametrization invariance of $S_U$. Then Eq. (\ref{59})
reduces to
\begin{equation}\label{66}
2i\dot\eta + (\dot U \g{_5}V)\eta=0
\end{equation}
and the substitution
\begin{equation}\label{67}
\eta = e^{\L(\t,\s)}\eta_{0}(\s),\: \,  (\eta_{0})^2 =0
\end{equation}
transforms  (\ref{66}) into the equation
\begin{equation}\label{68}
2i\dot \L+ (\dot U \g{_5}V)=0.
\end{equation}
In the gauges  (\ref{60}) and (\ref{65}) Eqs. (\ref{56})-(\ref{58}) take the
form
\begin{equation}\label{69}
U^a Y^\prime_{ab}U^b- 16 i\eta \eta^\prime =0,
\end{equation}
\begin{equation}\label{70}
\dot Y_{ab} U_b=0,
\end{equation}
\begin{equation}\label{71}
(\r^{\t} U_{a} U{_b})^{\cdot} =0,
\end{equation}
where the relation
\begin{equation}\label{72}
\eta\dot\eta =0
\end{equation}
resulting from (\ref{67}) has been  used.
Eq. (\ref{71}) multiplied by $(\g{_5}V)^a$ gives
 \begin{equation}\label{73}
\r^{\t}\dot U^{a}=[-\dot\r^{\t}+
\frac{1}{2i}\r^{\t}(\dot U \g{_5} V)]U^a
\end{equation}
and the equation (\ref{73}) can be written as
\begin{equation}\label{74}
\r^{\t}\dot U^{a}=-[\dot\r^{\t}+ \r^{\t}\dot \L]U^{a}
\end{equation}
after using  Eq.(\ref{68}).
The general solution of Eq.
(\ref{74}) is
\begin{equation}\label{75}
U^{a}(\t,\s)= \frac{1}{\r^{\t}}e^{-\L(\t,\,\s)}U^{a}_{0}(\s).
\end{equation}

The substitution of (\ref{75}) into  Eqs. (\ref{68}) and (\ref{71}) gives
\begin{equation}\label{76}
2\dot\L +  \frac{\dot\r^\t}{\r^\t}=0
\end{equation}
and  its  solution is
\begin{equation}\label{77}
\L(\t,\s)=-\frac{1}{2}\,ln\,[\r^{\t}\l_{0}(\s)],
\end{equation}
where $\l_{0}(\s)$ is an arbitrary function.

Using  (\ref{77}) one can present the general solutions of Eqs.
(\ref{66}) and (\ref{74}) as
\begin{eqnarray}\label{78}
\eta(\t,\s)= \frac{1}{\sqrt{\r^\t}}\,\eta_0(\s),\nonumber \\
U^{a}(\t,\s)= \frac{1}{\sqrt{\r^\t}}\, U^{a}_{0}(\s),
\end{eqnarray}
where $\l_{0}(\s)$ is contained in the redefinition of $\r^\t$ and  $U^{a}_{0}(\s)$.
The substitution of $ U^{a}$ (\ref{78}) into (\ref{13}) and of $\eta$ into
(\ref{64})
gives the following solutions  for $V^{a}$ and  $\th_a$
\begin{eqnarray}\label{79}
V^{a}(\t,\s)= \sqrt{\r^\t}\, V^{a}_{0}(\s),\nonumber \\
\th_a \equiv \th_a(\s)=-\eta_0(\s)(\g_{5}V_{0}(\s))_a
\end{eqnarray}
together with the constraint (\ref{13}) for the initial data $ U^{a}_{0}$ and
$V^{a}_{0}$
\begin{equation}\label{80}
 (U_0(\s)\g{_5}V_0(\s))=-2i.
\end{equation}
Note, that the $\t-$dependence of the dyads $U^{a}$ and $V^{a}$ can be
removed by fixing the residual reparametrization symmetry by the choice
$\r^\t(\t,\s)= \r^{\t}_0(\s)$ or $\r^\t(\t,\s)=const$. As a result, 
$U^{a}$ and
$V^{a}$ become  integrals of motion and will  be defined  by the initial
data of
the variational problem in question.
The representation (\ref{79}) shows that the Goldstone fermion 
corresponding to  partial spontaneous breaking of the $D=4,\, N=1$ global
supersymmetry describes a static spinor configuration distributed along the
closed superstring.

In view of (\ref{78}-\ref{79}) the remaining equations (\ref{70}-\ref{69})
take
the form
\begin{equation}\label{81}
(Y_{ab}U{_0}^b)^{\cdot}=0,
\end{equation}
\begin{equation}\label{82}
U_{0}^a Y^\prime_{ab}U_{0}^b-
16 i\eta_{0}\eta^\prime_{0} =0.
\end{equation}
Eq. (\ref{81}) shows that the Majorana spinor $Y_a$ defined by the relation
\begin{equation}\label{83}
i(\g_{5}Y)_a
\equiv Y_{ab}U_{0}^b,
\end{equation}
does not depend on the evolution parameter $\t$. We have used
$\g_{5}$ in the definition of $Y_a$ (\ref{83}) to present it  in the
canonical form
given by the Weyl representation (\ref{13})
\begin{equation}\label{84}
Y_{a}=
Y_{0a}(\s)\equiv
\left(y_{0\a}(\s) \atop \bar y^{\dot\a}_0(\s)\right).
\end{equation}
Then Eq. (\ref{82}) can be presented  in the form of the $\t$
independent constraint
\begin{equation}\label{85}
i[(U_{0} \g_{5} Y^\prime_{0}) - (U^\prime_{0} \g_{5} Y_{0})]-
16 i\eta_{0}\eta^\prime_{0} =0
\end{equation}
which  is  easily solved with respect to  $Y_{0}(\s)$ using the
expansion of $
Y^\prime_{0a}(\s)$ and $U^\prime_{0a}(\s)$ in the dyad basis as it is 
given by
the formula (\ref{17}). 
So, the solution of the constraint (\ref{85}) will
fix
the spinor $ Y_{0a}(\s)$ as a function of the initial data $U_{0a}(\s)$,
$V_{0a}(\s)$ and $U^\prime_{0a}(\s)$ 
fixed by the statement of the
variational problem under discussion. Note that the
initial
data  $U^\prime_{0a}$ are equivalent to the specification of the initial
data for
the derivative coefficients in  the $U^\prime_{0a}(\s)$ expansion in the dyad
basis. Therefore, one can consider the spinor $ Y_{0a}(\s)$ as given.
Then Eqs. (\ref{83}) are to be treated as equations for the restoration  of the
symmetric spinor matrix $Y_{ab}(\t,\s)$
\begin{equation}\label{86}
Y_{ab}(\t,\s)U{_0}^b(\s)= i(\g_{5} Y_{0}(\s))_a ,
\end{equation}
as a function of the given initial data $U_{0a}(\s)$ and $V_{0a}(\s)$.

The general solution of Eq.(\ref{86}) is presented by the sum
\begin{equation}\label{87}
Y_{ab}(\t,\s)=Y^{(inho)}_{ab}(\s)+Y^{(ho)}_{ab}(\t,\s),
\end{equation}
where $Y^{(inho)}_{ab}(\s)$ is the solution of the inhomogenous equation
\begin{eqnarray}\label{88}
Y^{(inho)}_{ab}(\s)=
\frac {i}{(U_0 \g_5 Y_{0})}
(\g_5 Y_0(\s))_{a}(\g_5 Y_{0}(\s))_{b}
\end{eqnarray}
and $Y^{(ho)}_{ab}(\t,\s)$ has to be the general solution of the homogeneous
equation
\begin{equation}\label{89}
Y^{(ho)}_{ab}(\t,\s)U_0^b(\s)=0.
\end{equation}
So, we see that the total $\t-$dependence of the spin-tensor $Y_{ab}(\t,\s)$
describing the superstring evolution in the enlarged superspace is
concentrated
in  $Y^{(ho)}_{ab}(\t,\s)$ and  the static solution
$Y^{(inho)}_{oab}(\s)$ may be treated as a domen-like vacuum solution.

To solve the homogeneous equation (\ref{89}) note that the relations
\begin{equation}\label{90}
\chi^\L_a(\t,\s)U_0^a(\s)=0,\:\, (\L=1,2,3)
\end{equation}
are satisfied by  the Majorana spinors $\chi^\L_a(\t,\s)$
\begin{equation}\label{91}
\chi^\L_a(\t,\s) \equiv \{U_a, \, (\g_5 U_a), \, V_a \}
\end{equation}
as it follows from the dyad definiton (\ref{12}-\ref{13}) and the solutions
(\ref{78}-\ref{79}).

Taking into account the relations (\ref{90}) we find the general solution of
Eq.(\ref{89})
\begin{eqnarray}\label{92}
Y^{(ho)}_{ab}(\t,\s)=\sum_{\L,\S}F_{\L\S}(\t,\s)\,
[ \chi^\L_{0a}(\s)\chi^\S_{0b}(\s) + \chi^\L_{0b}(\s)\chi^\S_{0a}(\s)],
\end{eqnarray}
where the $\t-$dependent factors $\sqrt{\r^\t}$ and $1/\sqrt{\r^\t}$
contained
in  $\chi^\L_a(\t,\s)$ have been  removed to redefine the arbitrary functions
$F_{\L\S}(\t,\s)$ parametrizing $Y^{(ho)}_{ab}(\t,\s)$.

Therefore, the general solution for the generalized coordinates
$Y_{ab}(\t,\s)$
(\ref{5}) describing superstring evolution in the superspace enlarged by TCC
coordinates is presented in the form
\begin{eqnarray}\label{93}
Y_{ab}(\t,\s)=
\frac {i}{(U{_0}(\s) \g_5 Y_{0}(\s))}(\g_5 Y_0(\s))_{a}(\g_5
Y_{0}(\s))_{b}\nonumber \\
+ \sum_{\L,\S}F_{\L\S}(\t,\s)\,
[\chi^\L_{0a}(\s)\chi^\S_{0b}(\s) + \chi^\L_{0b}(\s)\chi^\S_{0a}(\s)]
\end{eqnarray}
 with the total $\t-$dependence concentrated in the arbitrary scalar
coefficients $F_{\L\S}(\t,\s)$. This means that the superstring has no
transverse oscillations, but generates longitudinal exitations propagating in
the directions prescribed by the initial data for the dyads.

\section{Superbranes with $3/4$ $k-$symmetry}

The superstring action (\ref{54}) is naturally generalized to the case of
super
p-brane by means of enlargement of the range of values for the world-sheet
index $\mu$
\begin{eqnarray}\label{94}
S_p=\frac{1}{2}\int d\t d\s_1...d\s_p\,\r^\mu (U_a W_\mu^{ab} U_b),\:\:\:
\mu=(0,M),\: \: M=(1,2,..,p) ,
\end{eqnarray}
where $\r^\mu(\t,\s_M)$ is now a $k-$invariant world-volume density providing
  reparametrization invariance for  the p-brane action $S_p$. The
superstring
 equations of motion  (\ref{56}-\ref{59}) will preserve their form (modulo
the
$\mu$ dimesionality extension) and can be exactly solved using the same
method
as for the two-dimensional equations (\ref{56}-\ref{59}). To solve the p-brane
equations we choose the fermionic gauge condition (\ref{60}) and extend
 the bosonic gauge condition (\ref{65}) to
\begin{equation}\label{95}
\r^M(\t,\s_M)=0, \:\: M=(1,2,..,p).
\end{equation}
Then, we get  generalizations of the solutions (\ref{78}-\ref{79})
\begin{eqnarray}\label{96}
\eta(\t,\s_M)= \frac{1}{\sqrt{\r^\t}}\,\eta_0(\s_M),\nonumber \\
U^{a}(\t,\s_M)= \frac{1}{\sqrt{\r^\t}}\, U^{a}_{0}(\s_M), \nonumber \\
V^{a}(\t,\s_M)= \sqrt{\r^\t}\, V^{a}_{0}(\s_M),\nonumber \\
\th_a \equiv \th_a(\s_M)=-\eta_0(\s_M)(\g_{5}V_{0}(\s_M))_a.
\end{eqnarray}
 and change  the two remaining equations (\ref{81}-\ref{82}) by  the
following
system of $(p+1)$ equations
\begin{equation}\label{97}
(Y_{ab}U_{0}^b)^{\cdot}=0,
\end{equation}
\begin{equation}\label{98}
U_{0}^a \partial_M Y_{ab}U_{0}^b-
16 i\eta_{0} \partial_M\eta_{0} =0.
\end{equation}
The substitution (\ref{83}) used to introduce the Majorana spinor $Y_a$
can be
repeated again resulting in  the $\t-$independence of $Y_a$
\begin{equation}\label{99}
Y_{a}=
Y_{0a}(\s_M))\equiv
\left(y_{0\a}(\s_M) \atop \bar y^{\dot\a }_0(\s_M)\right).
\end{equation}
So, Eqs. $(\ref{98})$ are transformed into the $\t-$independent system
consisting of  $p$ constraints
\begin{equation}\label{100}
i[(U_{0} \g_{5}\partial_M Y_{0}) - (\partial_M U_{0} \g_{5} Y_{0})]-
16 i\eta_{0}\partial_M \eta_{0} = 0
\end{equation}
which generalize  the constraint (\ref{85}). These constraints can be solved
with respect to  $Y_{0}(\s_M)$ using the expansion (\ref{17}) for $
\partial_M Y_{0a}(\s_N)$ and $\partial_M U_{0a}(\s_N)$ in the dyad basis.
 This solution is to be completed by investigation of the
integrability conditions of these constraints. Modulo the integrability
conditions the solution for the generalized coordinates $Y_{ab}(\t,\s_M)$
(\ref{5}) is presented in the form (\ref{93}) with $\s_M$ substituted
instead of
$\s$. Note that
 super p-branes with the dimensions  p=3=dim$(x\,subspace)-1$,
p=5=dim$(z\,subspace)-1$ and  p=9=dim$((x \bigoplus z)\,space)-1$ 
may be a matter of  additional interest.
The solution (\ref{96}) for $\th_a(\s_M)$ presents the Goldstone fermion
corresponding to the partial spontaneous breaking of the $D=4,\ N=1$ global
supersymmetry and describes a static spin (or magnetic) configuration
distributed along the closed  p-brane. This solution can be treated as a
p-dimensional generalization of the magnetic Nielsen-Olesen vortex
discussed in
the previous
section.

\section{Conclusion}

We have considered a generalization of  twistor-like approach to describe
 superstrings and super p-branes evolved in the superspace enlarged by the
coordinates corresponding to the tensor central charges of the $D=4,\; N=1$
global supersymmetry algebra.
A simple class of supersymmetric and $\k-$symmetric actions preserving 3/4 of
the global supersymmetry has been proposed to prove the effect of the
supersymmetry enhancement in the string/brane dynamics as a result of the inclusion of the  TCC
coordinates. These superstring (\ref{51}) and super p-brane
(\ref{94})
actions are linear with respect to the diagonal matrix element  $W_U$
(\ref{29})
of the supersymmetric Cartan  one-forms $W_{ab}$ (\ref{9}) and generalize the
actions of tensionless strings and branes.
We have shown that the  requirement for $W_U$ to be an invariant of the
$\k-$symmetry imposes one real condition on the $\k-$symmetry parameter.
Alternatively, one could start choosing  $W_{\g_5 \vp}$ to be invariant,
but it
will not change the conclusions.
The requirement for another diagonal element to be an additional invariant
 of the $\k-$symmetry imposes one more real condition reducing the number of
preserving supersymmetries to two instead of three ones.
As a result, a super p-brane action including the diagonal elements $W_U$ and
$W_V$, e.g. the quadratic superstring action
\begin{equation}\label{101}
S_{(U,V)}=g\int d\t d\s \,\e^{\mu\nu}(U_a W_\mu^{ab} U_b)(V_a W_\nu^{ab}
V_b),
\end{equation}
 will preserve 1/2 of the global supersymmetry. To restore the lost extra
$\k-$symmetry it is necessary to add some compensating terms such as the
Wess-Zumino terms or other differential 2-forms constructed of the total
set of
the Cartan forms and/or their matrix elements. We intend to discuss this
problem
in another place.

A physical content of the studied models with enhanced  $\k-$symmetry can be
understood attracting the ideas advanced in \cite{hupol},\cite{hulpol} and
based
on the conceptions developed in \cite{vol},\cite{no}. Note also that in
\cite{zgt} string in the D=4 space-time is  described by a closed
exactly
solvable sector of the $SO(1,1)$x$O(2)$ two-dimensional gauge model 
connected with the underlying gauge theory for the Nielsen-Olesen vortex.

We have shown that the presence of extra $\k-$symmetry in the proposed linear
actions allows to find the general solutions for the Goldstone fermion and
bosons corresponding to the partial spontaneous breaking of the global
supersymmetry.
We can try to use the Goldstone fermionic solution to describe a magnetic
configuration associated with the Nielsen-Olesen vortex or with  its p-dimensional
generalization described by the long wave p-brane approximation. So,
despite the
fact that the fermions associated with supersymmetry do not carry electric
charge they may  describe magnetic properties of superstring and
superbranes.
This  property of the Grassmannian coordinates $\th_a$ was revealed in
\cite{tz1},\cite{tz2}, where a supersymmetric generalization of the
Fokker-Schwarzschild-Tetrode-Wheeler-Feynman electromagnetic theory was
studied.
It was shown there that the spinors $\th_a$ contribute to
an anomalous magnetic moment (AMM) of neutral particles with spin 1/2.
Also, it was shown in \cite{uz} that the contribution of the fixed AMM of D=4
N=2 charged massive superparticle interacting with extended Maxwell
supermultiplet restores $\k-$symmetry broken by the minimal coupling
\cite{lm}.

Realization of such a possibility is supported by the following hint.
As follows from the solution of the superstring and superbrane
equations of
motion, the effective spinors $Y_a$ (\ref{84}) and (\ref{99}) are the first
integrals
\begin{equation}\label{102}
 \dot Y_a=0
\end{equation}
of these equations. Using this result one can propose to consider the preserving bilinear covariant 
 \begin{equation}\label{103}
 J_{MN}=i Y_a (C\S_{MN})^a{}_b Y^b, \:\: \: \dot J_{MN}=0,
\end{equation}
where $\S_{MN}$ is the Lorentz group generators in the Majorana
representation, to describe  an effective spin/orbital  momentum density of
superstring/superbrane. Then the gauge invariant Lagrangian density
\begin{equation}\label{104}
L_{e.m.}=\mu^* F^{MN}J_{MN},
\end{equation}
may be treated as an analog of electromagnetc interaction of spin 1/2  particles by means of their AMM, if $\mu^*$ has a sense of the phenomenological constant describing the value of the effective AMM of the superstring/superbrane.

Using the proposal \cite{hupol} to consider cosmic supersymmetric
Nielsen-Olesen
vortices, it is interesting to treat the integrated AMM density of the vortex
as a source of strong magnetic fields associated with cosmological objects
and
dark matter. Here  we use an analogy  with atomic spins of a
crystal 
lattice as sources of magnetic fields in the  domain walls of 
ferromagnetics.
 The Lagrangian of spin waves treated as the Goldstone particles in magnetic
media
was constructed in \cite{dvzbl},\cite{dvzsw}. Applying this  approach one
can try
to consider the Goldstone solutions associated with the p-vortices as
standing
spin waves distributed along the closed superstring or super p-brane.

Of course, fixing exact physical contents of the considered model supposes
carrying out its covariant quantization. According to the results which have been obtained here a suitable  effective variable for the quantization in the fixed gauge (\ref{60}) is the Majorana bispinor $Y_a$ (see (\ref{83}))
\begin{equation}\label{105}
 i(\g_{5}Y)_a
\equiv Y_{ab}U_{}^b,
\end{equation}
and the initial data $Y_{0a}(\s_M)$ and $U_{0a}(\s_M)$ appear to be proper primary variables for the canonical quantization in view of the solution (\ref{93}).
In terms of $Y_a$ and  $\eta$ (\ref{61}) the superstring  (\ref{54}) and
super p-brane (\ref{94})  actions are equivalenly presented as
\begin{eqnarray}\label{106}
S_p=\frac{i}{2}\int d\t d\s_1...d\s_p\,\r^\mu \{[(U\g_{5}\partial_\mu Y)
-(\partial_\mu U\g_{5} Y)] -
\eta \partial_{\mu}\eta +
2i\partial_\mu U^{a}\eta \th_a \}.
\end{eqnarray}
The last term $\partial_\mu U^{a}\eta \th_a$ in (\ref{106}) is gauge dependent and it  vanishes in the gauge (\ref{60}). The restoration of the $\eta \th_a$ term  in the definition of $Y_a$ (\ref{105}) yields the new effective variable $\tilde Y_a$
\begin{equation}\label{107}
 (\g_{5}Y)_a= \tilde Y_a - i \eta \th_a. 
\end{equation}
The substitution of the shifted field $\tilde Y_a $ in (\ref{106}) results in the new representation 
\begin{eqnarray}\label{108} 
S_p=\frac{i}{2}\int d\t d\s_1...d\s_p\,\r^\mu \{[(U^a\partial_\mu \tilde Y_a)
-(\partial_\mu U^a \tilde Y_a)] - \eta \partial_{\mu}\eta \}
\end{eqnarray}
which includes only the $\k-$invariant Goldstone fermion $\eta$ representing the whole fermionic sector of the super p-brane. The bosonic sector, originally presented by the primary bosonic variables $Y_{ab}$, or equivalently by  $x_{m}$ and $z_{mn}$, is encoded by $\tilde Y_a$. 
So, we conclude that the transformation of variables 
\begin{equation}\label{109}
 Y_{ab}U_{}^b = i\tilde Y_a + \frac{1}{2i}(U^b \th_b) \th_a
\end{equation}
eliminates the correspondent gauge degree of freedom from the original actions
(\ref{54}) and  (\ref{94}) without using any gauge conditions.  
  It shows a universal property of the extra $ \k-$symmetry to remove not only the fermionic gauge  degrees of freedom, but also the bosonic ones originally introduced by the coordinates $x_m $ and $z_{mn}$. 

The representation  (\ref{108}) generalizes the supertwistor formulations of superparticle
action \cite{balu}, \cite{fer}, \cite{shir}, \cite{bebcel}, \cite{luknow}  to the case of superstrings and superbranes, and includes eight bosonic fields described by the Majorana spinors $U_a(\t,\s^M)$, $\tilde Y _a(\t,\s^M)$  and one fermionic field described by the  Grassmann variable $\eta(\t,\s^M)$ which can be treated as the components of a real supertwistor. 
The property of the formulation (\ref{108}) to include the gauge invariant  variables (in the reparametrization gauge (\ref{95}) and $\r^0(\t,\s^M)=const$ ) suggests  a realization of the gauge independent quantization in terms of the supertwistor components $(U_a,\tilde Y^a, \eta)$. 
In the twistor description of massless superparticle (see, e.g. \cite{shir}) this is a reason to consider the space of twistors as more fundamental than space-time. May be that point of view will give some 
advantages in the super p-brane theory.
In any case, the quantization in terms of the supertwistor $(U_a,\tilde Y^a, \eta)$ have to be an effective way  to find the spectrum of the super p-brane model with enhanced supersymmetry. Study of this problem is in progress now.
 
\section {Acknowledgements}

A.Z. would like to acknowledge  Fysikum at the Stockholm University for kind
hospitality. He is also grateful to Ingemar Bengtsson, Lars Bergstr\"om, Ulf
Danielsson, Ulf Lindstr\"om and Hector Rubinstein for fruitful
discussions. The
work is partially supported by the grant of the Royal Swedish Academy of
Sciences and Ukrainian SFFR project 02.07/276.
A.Z. was  partially supported by the grant of Axel Wenner-Gren Foundation
 and the Award CRDF-RP1-2108.

\end{document}